# AI-Driven Acoustic Voice Biomarker-Based Hierarchical Classification of Benign Laryngeal Voice Disorders from Sustained Vowels

Mohsen Annabestani[1], Samira Aghadoost[2], Anaïs Rameau[3], Olivier Elemento[4], Gloria Chia-Yi Chiang[5]

1.  Dalio Institute of Cardiovascular Imaging, Department of Radiology, Weill Cornell Medicine
2.  Department of Speech Therapy, School of Rehabilitation, Tehran University of Medical Sciences
3.  Sean Parker Institute for the Voice, Department of Otolaryngology-Head & Neck Surgery, Weill Cornell Medicine
4.  Englander Institute for Precision Medicine, Department of Physiology and Biophysics, Weill Cornell Medicine
5.  Brain Health Imaging Institute, Department of Radiology, Weill Cornell Medicine

**Abstract**

Benign laryngeal voice disorders affect nearly one in five individuals and often manifest as dysphonia, while also serving as non-invasive indicators of broader physiological dysfunction. We introduce a clinically inspired hierarchical machine learning framework for automated classification of eight benign voice disorders alongside healthy controls, using acoustic features extracted from short, sustained vowel phonations. Experiments utilized 15,132 recordings from 1,261 speakers in the Saarbruecken Voice Database, covering vowels /a/, /i/, and /u/ at neutral, high, low, and gliding pitches. Mirroring clinical triage workflows, the framework operates in three sequential stages: Stage 1 performs binary screening of pathological versus non-pathological voices by integrating convolutional neural network–derived mel-spectrogram features with 21 interpretable acoustic biomarkers; Stage 2 stratifies voices into Healthy, Functional or Psychogenic, and Structural or Inflammatory groups using a cubic support vector machine; Stage 3 achieves fine-grained classification by incorporating probabilistic outputs from prior stages, improving discrimination of structural and inflammatory disorders relative to functional conditions. The proposed system consistently outperformed flat multi-class classifiers and pre-trained self-supervised models, including META HuBERT and Google HeAR, whose generic objectives are not optimized for sustained clinical phonation. By combining deep spectral representations with interpretable acoustic features, the framework enhances transparency and clinical alignment. These results highlight the potential of quantitative voice biomarkers as scalable, non-invasive tools for early screening, diagnostic triage, and longitudinal monitoring of vocal health.

**Keywords:** Laryngeal voice disorders, Hierarchical classification, HuBERT, HeAR, Mel-spectrograms, Digital Health, Vocal Biomarkers

## 1. Introduction

Voice is one of the most fundamental and complex instruments of human communication. It supports linguistic exchange while also transmitting emotion, identity, and social intent [1, 2]. Maintaining a healthy and stable voice is essential for effective interpersonal interaction and for professional performance, particularly in vocally intensive fields such as teaching, acting, singing, and broadcasting [3-5]. The ability to produce a clear and fatigue-resistant voice enables individuals to communicate efficiently, sustain social relationships, and meet the demands of their professional roles [6, 7]. In contrast, disruptions in vocal function, collectively described as voice disorders, can substantially hinder social participation, emotional expression, and occupational capabilities [8, 9].

Voice disorders affect an estimated 7-10% of individuals worldwide at some point in their lives, highlighting their clinical and public health significance [10]. Their etiologies are diverse and include functional and psychogenic origins such as hyperfunctional dysphonia, muscle tension dysphonia (MTD), and vocal misuse. They also include structural and pathological conditions such as vocal fold nodules, polyps, Reinke's edema, laryngitis, and pachydermia [11, 12]. Environmental exposures including air pollution and occupational vocal load, lifestyle factors including smoking and caffeine intake, and systemic diseases such as gastroesophageal reflux and hypothyroidism further contribute to the development and persistence of these disorders [12, 13].

Among their manifestations, dysphonia is the most prevalent and clinically quantifiable symptom. It is characterized by abnormalities in pitch, loudness, and timbre, and is perceived as hoarseness, breathiness, or roughness [14, 15]. Persistent dysphonia reduces vocal endurance, compromises professional performance, and negatively affects overall quality of life [16, 17]. It may also obscure early signs of more severe conditions such as vocal fold paralysis or laryngeal carcinoma. For these reasons, timely and accurate differentiation of underlying etiologies is essential for selecting appropriate behavioral, medical, or surgical treatment strategies [18, 19].

Conventional diagnostic approaches combine auditory-perceptual evaluation, acoustic analysis, laryngoscopic imaging, palpation, and patient-reported measures [20, 21]. Some methods, such as stroboscopy, require specialized equipment, trained clinicians, and in-person assessment. These requirements introduce cost, invasiveness, and logistical barriers that restrict accessibility, particularly in remote or resource-limited settings. During global health crises such as the COVID-19 pandemic, the risks associated with in-person evaluations further emphasized the need for scalable, non-invasive, and remote diagnostic alternatives [22]. Acoustic analysis provides an objective and reproducible option by quantifying features such as jitter, shimmer, fundamental frequency (F0), and harmonic-to-noise ratio (HNR) [23, 24]. Sustained vowel phonation tasks, including /a/, /i/, and /u/, are especially advantageous because they are language independent, easy to elicit, and sensitive to subtle perturbations in vocal fold vibration [25]. However, manual acoustic assessment is time-consuming and reliant on expert interpretation, which may result in inter-rater variability and diagnostic inconsistency.

Recent machine and deep learning advances have greatly improved automated classification of voice disorders. CNNs, LSTMs, BiLSTMs, CNN–RNN hybrids, and ensemble methods use acoustic inputs such as MFCCs, spectrograms, Mel-spectrograms, TQWT features, and glottal parameters to capture key spectral and temporal cues linked to dysphonia[26-29]. Pretrained and custom CNNs applied to spectrograms from sustained vowels often achieve strong results for binary pathology detection, while BiLSTMs trained on MFCC sequences from continuous speech provide better modeling of temporal patterns[26]. Hybrid systems that pair CNN feature extraction with LSTM sequence modeling, as well as ensembles based on EfficientNet, ResNet, or DenseNet, further improve performance. Gradient boosting models such as XGBoost combined with handcrafted features like jitter, shimmer, and HNR also remain competitive[26, 28, 29].

Model performance is strongly influenced by the dataset used. The Saarbruecken Voice Database (SVD) [30] is a common benchmark for sustained-vowel classification, while clinical datasets and resources such as VOICED or Mandarin continuous-speech corpora support evaluations with more complex speech [26, 27, 31]. Binary healthy-versus-pathology classification often reports accuracies above 90%. In contrast, multi-class subtype classification shows lower unweighted average recalls, typically in the range of 50% to 70% [26, 28, 31] even with Large Speech Models like HuBERT[32]. Although some CNN-based models achieve up to 95 percent accuracy or F1-scores near 94 percent on SVD subsets [33], cross-database experiments reveal substantial performance drops. For example, XGBoost with acoustic and MFCC fusion reaches an F1 of 0.733, compared to 0.621 for DenseNet [26, 29]. This wide performance range, often between 70 and 99 percent, underscores the need for consistent dataset splits, demographic control, and standardized evaluation practices.

The Saarbruecken Voice Database (SVD) [30] is a widely adopted benchmark dataset in machine learning-based voice pathology research, serving as a foundational resource for numerous studies developing automated detection and classification models [32-38]. However, comparisons across studies remain challenging due to inherent variations in the SVD's pathology subsets, as the dataset encompasses a broad spectrum of disorder types from which researchers typically select only specific subsets for analysis. Consequently, models are evaluated on disparate disorder combinations, rendering direct comparisons of reported accuracies unreliable. Furthermore, the majority of these efforts prioritize binary classification tasks [35-38], with model performance highly contingent on the chosen data subset and vowel selections (e.g., /a/, /i/, or /u/ phonations), which further exacerbates inconsistencies in cross-study benchmarking.

Clinically, these models support noninvasive screening, remote monitoring, and rapid triage. Lightweight CNN architectures allow deployment on mobile or edge devices for practical use in healthcare settings[28]. Despite progress enabled by multi-feature fusion and continuous-speech modeling, challenges remain in interpretability, external validation, and fairness across different demographic or linguistic groups. Future directions include explainable modeling, stronger cross-database validation, and broader adoption of transformer-based and self-supervised learning approaches to improve clinical readiness.

## 2. Materials and Methods

### 2.1 Dataset and Experimental Design

To conduct this study, we utilized a subset of the Saarbrücken Voice Database (SVD)[30], a widely recognized and clinically curated repository of vocal recordings extensively used in voice pathology research. From the full database of voice recordings, we selected 15,132 sustained vowel samples from 1,261 participants, ranging in age from 18 to 85 years. The dataset is slightly imbalanced, with a female majority of about 10% (~60% female compared to ~40% male).

Each participant was instructed to produce 12 recordings. This included the vowels /a/, /i/, and /u/ in four pitch conditions: neutral, high, low, and a continuous low to high to low gliding intonation. These recordings form the dataset used for training our classifier.

The samples were grouped into nine categories, including one healthy control group and eight distinct vocal pathology subtypes, as shown in **Table 1**. In total, 54.0% of participants were classified as Non-Pathological (Healthy) and 46.0% as Pathological. The gender distribution in both groups showed a similar pattern, with a slight female majority. The Healthy group consisted of 62.0% women and 38.0% men, and the Pathological group consisted of 59.3% women and 40.7% men. With its controlled recording conditions, clinically validated labels, and high quality audio, this dataset provides a reliable foundation for developing machine learning models for detecting and classifying voice disorders.

The disorders included in this study represent the core spectrum of benign laryngeal voice disorders, which traditionally span structural, inflammatory, and functional/psychogenic etiologies. For analytical clarity and clinical triage, our work consolidates these into two broader groups:

- **Structural and Inflammatory Disorders:** These capture the essential organic pathologies that alter vocal fold mass, tissue properties, and mucosal vibration. The included disorders are *Laryngitis, Contact Pachydermia, Reinke's Edema*, and *Vocal Cord Polyp*.
- **Functional and Psychogenic Disorders:** These reflect maladaptive or non-organic mechanisms that disrupt normal phonation without overt structural lesions. The included disorders are *Functional Dysphonia, Hyperfunctional Dysphonia, Psychogenic Dysphonia,* and *Dysodia*.

Together, these conditions constitute the clinically central and most frequently encountered benign laryngeal disorders. To ensure generalizability and prevent data leakage, the dataset was subjected to a subject-independent partition, separating recordings by individual participant. The resulting split was 80% for training, 10% for validation, and 10% for final, independent testing.

**Table 1**. Baseline Cohort Demographics by Diagnosis and Sex. Distribution of the 1,261 unique participants, categorized by the nine vocal pathology groups (Non-Pathological and Pathological).

| Diagnosis (Pathologies) | Total Count | Female Count | Male Count | Female Proportion | Male Proportion |
|---|---|---|---|---|---|
| Hyperfunctional Dysphonia | 173 | 131 | 42 | 75.7% | 24.3% |
| Laryngitis | 117 | 48 | 69 | 41.0% | 59.0% |
| Functional Dysphonia | 92 | 62 | 30 | 67.4% | 32.6% |
| Dysodia | 51 | 35 | 16 | 68.6% | 31.4% |
| Psychogenic Dysphonia | 44 | 29 | 15 | 65.9% | 34.1% |
| Contact Pachydermia | 42 | 2 | 40 | 4.8% | 95.2% |
| Reinke Edema | 35 | 29 | 6 | 82.9% | 17.1% |
| Vocal Cord Polyp | 26 | 8 | 18 | 30.8% | 69.2% |
| **Total Pathological** | **580** | **344** | **236** | **59.3%** | **40.7%** |
| Healthy | 681 | 422 | 259 | 62.0% | 38.0% |
| **Total Non-Pathological** | **681** | **422** | **259** | **62.0%** | **38.0%** |
| **Overall Cohort Total** | **1261** | **766** | **495** | **60.7%** | **39.3%** |

## 2.2 Feature Extraction and Representation

The proposed framework employs two complementary feature sets to capitalize on both spectral patterns and established acoustic biomarkers.

### 2.2.1 Deep Spectral Features (Mel-Spectrograms)

In the first stage of our deep learning framework, we utilized two-dimensional Mel-spectrograms derived from sustained vowel recordings as the primary input features. These representations capture the time-frequency distribution of acoustic energy, enabling the identification of complex, nonlinear vocal fold irregularities, such as noise components and subharmonics, that characterize dysphonic speech. The acoustic signals were processed at a sampling rate of 44.1 kHz, applying a Fast Fourier Transform (FFT) with a window size of 1,024 samples and a hop length of 128 samples to achieve high temporal resolution. This process generated 128 Mel-frequency bands, which we converted into a logarithmic decibel scale. To ensure data uniformity for the Convolutional Neural Network, we statistically normalized each spectrogram and fixed the temporal dimension to 256-time steps using truncation or padding where necessary.

### 2.2.2 Handcrafted Acoustic Biomarkers

For the subsequent machine learning stages, a set of 21 traditional acoustic and demographic features, widely recognized in clinical phoniatrics as essential voice biomarkers, were extracted:
- **Demographic Feature:** Gender (encoded as a binary variable).
- **Perturbation and Noise Measures:** Shimmer (quantifying amplitude perturbation), and Harmonics-to-Noise Ratio (HNR), which inversely correlates with perceived roughness and breathiness.
- **Fundamental Frequency and Formants:** The fundamental frequency (F0), and the first three formants (F1, F2, F3), which are critical indicators of vocal fold vibration rate and vocal tract resonance characteristics.
- **Timbre Descriptors:** The first 13 Mel-Frequency Cepstral Coefficients (MFCCs), capturing spectral envelope information related to vocal quality.

## 2.3 The Hierarchical Classification Framework

Designed to replicate clinical triage, our classification architecture employs a three-stage hierarchical strategy progressing from general screening to specific diagnosis (**Figure 1**). The process begins with Stage 1, a binary screen distinguishing pathological from healthy samples. Stage 2 subsequently classifies cases into broad etiological categories; Healthy, Functional/Psychogenic, and Structural/Inflammatory. The framework concludes with Stage 3, which performs a granular analysis to identify the precise disorder from nine distinct subtypes.

### 2.3.1 Stage 1: Binary Screening (Pathological vs. Non-Pathological)

The primary objective of this initial phase is to function as a high-sensitivity screening tool capable of flagging any deviation from a healthy vocal state. We implemented this using a two-step feature fusion strategy:

- **Step 1:** We deployed a Convolutional Neural Network (CNN) to process the Mel-spectrogram data directly for binary classification. The network architecture comprises three sequential convolutional blocks. Each block contains a 2D convolutional layer, utilizing 32, 64, and 128 filters respectively with 3x3 kernels, followed by a Rectified Linear Unit (ReLU) activation function, Batch Normalization to stabilize learning, and a Max-Pooling layer to reduce spatial dimensions.
- **Step 2:** To leverage the strengths of both deep representation learning and acoustic theory, we extracted the softmax probability outputs from the CNN, treating them as high-level "probability features" representing the likelihood of Pathological versus Non-Pathological states. We concatenated these 2 deep-learning-derived values with the 21 traditional acoustic biomarkers (comprising 19 numeric features like MFCCs and Shimmer, plus encoded gender data) to create a consolidated feature vector of 23 dimensions.

### 2.3.2 Stage 2: Etiological Triage (3-Class)

This stage performs classification into the three clinical groups: Healthy, Functional/Psychogenic, and Structural/Inflammatory. To inform this decision, the Stage 1 final binary output (one-hot encoded, with a threshold of 0.5) was concatenated with the 21 handcrafted acoustic features, resulting in a 22-dimensional feature vector. The successful output of Stage 1 acts as a powerful prior diagnostic biomarker, significantly enhancing the model's ability to discriminate between the broad etiologies.

### 2.3.3 Stage 3: Subtype Classification (9-Class)

The final stage targets the precise classification among the nine specific disorder classes. The feature set for this stage was comprehensively augmented by concatenating the 21 handcrafted acoustic features with the output of Stage 1 (one feature) and the output of Stage 2 (three features), creating a robust 25-dimensional feature space. This ensures the fine-grained classifier benefits from the entire diagnostic history provided by the preceding stages.

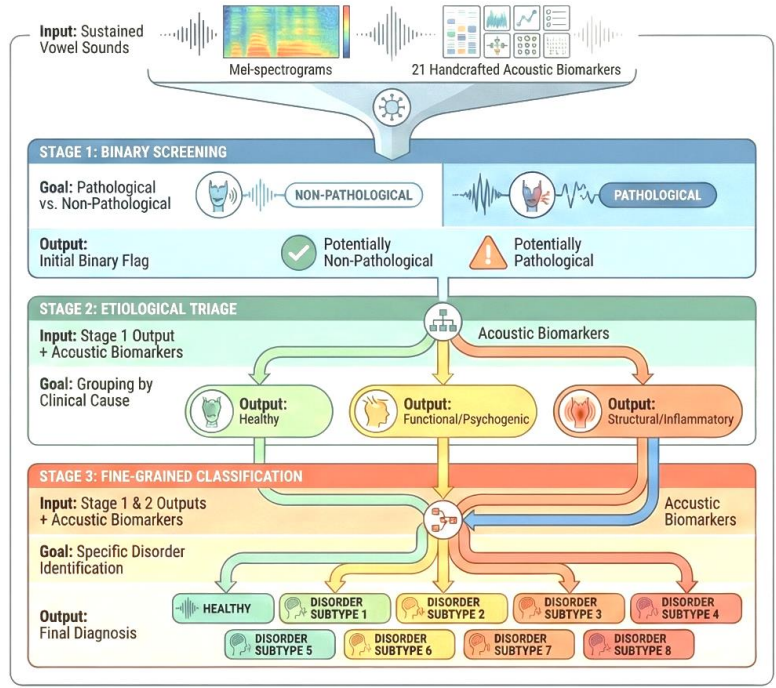

**Figure 1.** Proposed 3-Stages Machine Learning Hierarchical Classification Framework. *This figure was generated using the NotebookLM AI tool.*

## 2.4 Models and Training Details

For the first part of Stage 1, the CNN model, which was developed in Python, we used a standard method to refine its learning. We trained the CNN for 50 cycles (epochs) using the Adam optimizer and the cross-entropy loss function. We implemented an early stopping rule if the performance on a separate validation set didn't improve.

For the second part of Stage 1, as well as in all of Stages 2 and 3, we tested a large variety of machine learning models; over 20 in total. This diverse selection included common techniques like Decision Trees, Logistic Regression, Naive Bayes, Support Vector Machines (SVMs), various ensemble methods, and neural networks. We managed this process using the MATLAB Classification Learner. At the end of each stage, we simply selected the model that performed the best to make the final classification decision. To ensure the models were perfectly tuned, we relied on the learner's integrated Bayesian optimization to automatically find the best settings (hyperparameters). This optimization involved a robust 5-fold cross-validation process. Crucially, the final performance metrics for every stage were verified using a separate, reserved test set.

To evaluate the generalization performance of our hierarchical framework relative to pre-trained Large Speech Models (LSMs), we conducted comparative benchmarking against two representative self-supervised audio models. We fine-tuned META HuBERT (Hidden-Unit Bidirectional Encoder Representations from Transformers) [39] as a transformer-based LSM baseline and included Google HeAR (Health Acoustic Representations)[40] as a domain-relevant self-supervised audio embedding baseline. HuBERT is a transformer architecture trained via self-supervised learning to predict masked discrete acoustic units obtained through clustering of

MFCC-derived features. This learning paradigm enables HuBERT to acquire rich contextual representations of speech that have demonstrated strong transfer performance in downstream tasks such as automatic speech recognition. In contrast, HeAR is explicitly designed to learn health-relevant acoustic embeddings using a masked autoencoder framework trained on large-scale non-semantic health audio, including coughs, breathing, and other respiratory sounds.

For both models, audio recordings were resampled to 16 kHz, converted to mono, and either padded or truncated to a fixed duration of 3 seconds (48,000 samples) to ensure uniform inputs. For HeAR, audio was additionally transformed into mel-spectrograms with 192 mel bands and 128 time frames prior to embedding extraction. HuBERT embeddings were extracted from the final (12th) transformer layer, yielding 768-dimensional feature vectors that were classified using a linear classifier trained with the Adam optimizer, cross-entropy loss, and validation-based learning-rate scheduling. HeAR embeddings were evaluated using a support vector machine with a radial basis function (RBF) kernel, consistent with its intended use as a frozen, general-purpose health acoustic representation. This experimental setup enabled a direct and controlled comparison, revealing that while HuBERT performs well on general speech modeling, its conversational speech pre-training limits sensitivity to dysphonic and health-specific acoustic irregularities. Although HeAR is optimized for health acoustics, its generic pre-training objective without task-specific adaptation resulted in lower-than-expected performance for dysphonic analysis, further motivating the need for the domain-aware hierarchical modeling strategy proposed in our framework.

## 3. Results

To assess how well our hierarchical classification framework performs, we evaluated it using a broad set of metrics that capture both overall accuracy and its stability across the different classes at each stage. Because the system operates in three sequential classification steps, we relied on standard performance measures derived from the confusion matrix to quantify the model's behavior. The core metrics used to evaluate predictive performance are *Accuracy, Precision, Recall, and F1-Score*. These are computed from the standard confusion matrix terms: True Positives (TP), True Negatives (TN), False Positives (FP), and False Negatives (FN). We also report two additional measures: *ROC-AUC* (Receiver Operating Characteristic – Area Under the Curve) and *PR-AUC* (Precision–Recall – Area Under the Curve).

The main metrics are defined as follows:

**Accuracy:** Indicates how often the model's predictions were correct overall.

$$Accuracy = \frac{TP+TN}{TP+TN+FP+FN} \tag{1}$$

**Precision (Positive Predictive Value):** Reflects the proportion of predicted positive cases that were truly positive.

$$Precision = \frac{TP}{TP+FP} \tag{2}$$

**Recall (Sensitivity):** Shows the proportion of actual positive cases that the model successfully identified.

$$\text{Recall} = \frac{TP}{TP+FN} \quad (3)$$

**F1-Score:** The harmonic mean of Precision and Recall, offering a balanced assessment of performance, especially in multi-class settings.

$$\text{F1} - \text{Score} = 2 \times \frac{Precision \times Recall}{Precision+Recall} \quad (4)$$

Because the dataset is imbalanced across the eight pathology subtypes and the healthy control group, we report Precision, Recall, and F1-Score using two complementary averaging strategies. *Macro-Averaging* computes each metric independently for every class and then takes the simple average, which treats all classes equally regardless of size. *Micro-Averaging*, on the other hand, aggregates TP, FP, and FN across all classes before calculating the metric, effectively weighting the results by the number of samples and giving more influence on the majority classes. This combination of metrics provides a comprehensive view of how the model performs both within each classification stage and across the full hierarchy.

### 3.1 Stage 1: Refined Binary Screening Performance

The two-step feature-fusion strategy used for the binary classification task (Pathological vs. Non-Pathological) led to a clear performance enhancement. By combining the soft classes probability outputs from the CNN with the handcrafted acoustic features, the model gained access to both high-level learned representations and complementary signal-driven information. This enriched feature space allowed the final Stage 1 custom gaussian SVM classifier (as the best perform model in the pool, **Figure 2**) to make more robust decisions. As a result, the fused model reached a test accuracy of 80.5%, representing an improvement compared with the CNN-only baseline accuracy of 76.3%. Both the CNN-only and fused SVM models outperformed the HuBERT, and Google HeAR models, which achieved accuracy of 74.9% 58.43% respectively. A full breakdown of Precision, Recall, and F1-Score, for this stage is reported in **Table 2**.

**Table 2.** Aggregated and Per-Class Performance Metrics for Stage 1 (Binary Classification - Pathological vs. Non-Pathological) using the selected Gaussian SVM classifier.

|  | Metric | Precision | Recall | F1-Score |
|---|---|---|---|---|
| **Aggregated** | Macro | 80.5% | 79.5% | 79.7% |
|  | Micro | 80.0% | 80.0% | 80.0% |
| **Per-Class** | Non-Pathological | 77.8% | 74.0% | 75.9% |
|  | Pathological | 82.2% | 85.1% | 83.6% |

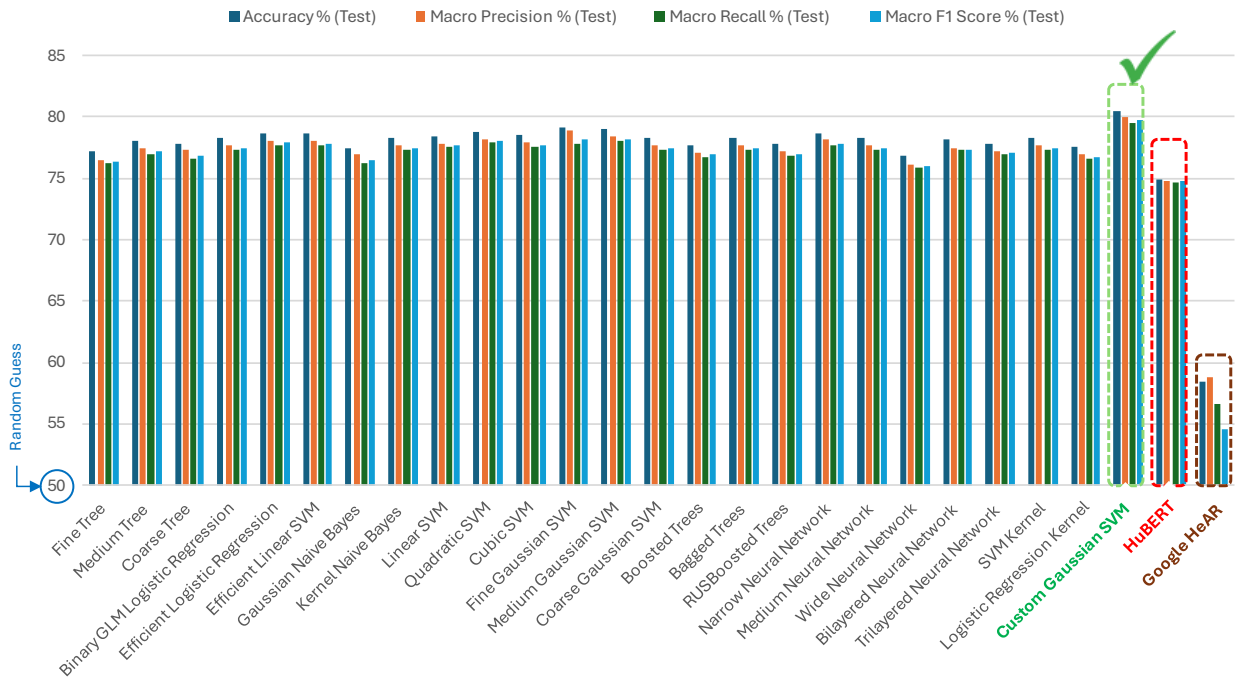

**Figure 2.** Test performance comparison across all evaluated models in Stage 1. The custom Gaussian SVM (Green) achieved the highest accuracy, outperforming all models including HuBERT (Red), and Google HeAR (Brown), and was selected as the best-performing classifier.

The model achieved a recall of 85.1 percent for the Pathological class, indicating that it successfully identified the majority of disordered voices and demonstrates its promise as a sensitive tool for early clinical screening. This level of sensitivity is particularly important in real-world diagnostic workflows, in which missing pathological cases can delay intervention and negatively affect patient outcomes. To provide a clearer picture of the model's performance across different decision thresholds, **Figure 3** presents the ROC and Precision–Recall curves, illustrating how the classifier balances sensitivity with specificity and how precision varies with recall. The Macro-averaged ROC-AUC of 0.848 and PR-AUC of 0.830 further attest to the model's strong discriminative ability and stability across classes. Complementing these visualizations, the confusion matrix in **Figure 4** details the distribution of correct and incorrect predictions, offering additional insight into class-specific performance patterns and potential areas for improvement.

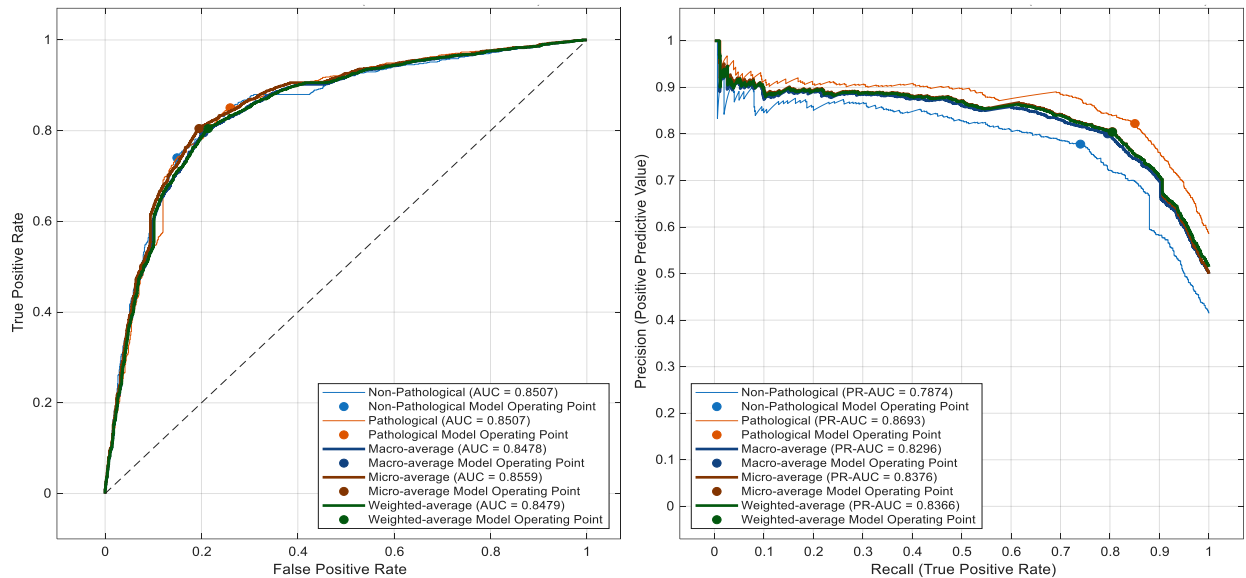

**Figure 3.** ROC and Precision-Recall Curves for Stage 1 Binary Classification using the selected Gaussian SVM classifier. This figure displays the Receiver Operating Characteristic (ROC) curve and the Precision-Recall (PR) curve for the refined binary classification model (Stage 1), illustrating the trade-off between sensitivity and specificity, and precision and recall, respectively. The Macro-averaged AUC scores (ROC-AUC: 0.848, PR-AUC: 0.830) confirm the robust discriminative power of the fused feature set.

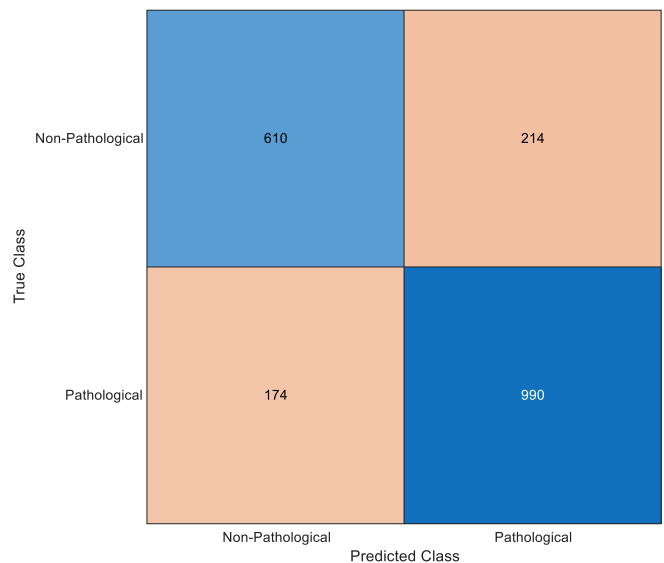

**Figure 4.** Confusion Matrix for Stage 1 Binary Classification using the selected Gaussian SVM classifier. The confusion matrix visually represents the performance of the Stage 1 classifier on the test set. It highlights the true positive, true negative, false positive, and false negative rates for Pathological and Non-Pathological classes, demonstrating the model's high recall for pathological samples.

### 3.2 Stage 2: Etiological Triage Performance

The goal of Stage 2 was to further categorize each sample into one of three clinically meaningful groups: Healthy, Functional/Psychogenic, or Structural/Inflammatory. To strengthen this multi-class decision, the binary output from Stage 1 was incorporated as an auxiliary biomarker, providing additional context regarding the presence or absence of pathology. Among the pool of more than twenty candidate models evaluated for this task, the cubic SVM emerged as the top performer, achieving an overall accuracy of 86.7 percent (**Figure 5**). This model consistently outperformed alternatives across key evaluation metrics. The cubic SVM models also outperformed the HuBERT model, which achieved an accuracy of 62.4%. A detailed summary of the Precision, Recall, and F1-Score for the aggregated and per-class performance is provided in **Table 3**.

**Table 3.** Aggregated and Per-Class Performance Metrics for Stage 2 (three-classes Classification - Healthy, Functional/Psychogenic, Structural/Inflammatory) using the selected cubic SVM classifier. The perfect score for the Healthy class (100%) reflects the deliberate architecture of our system, where the output of the Stage 1 binary classifier effectively pre-separates the Non-Pathological (Healthy) samples.

|  | Metric | Precision | Recall | F1-Score |
|---|---|---|---|---|
| **Aggregated** | Macro | 83.0% | 82.7% | 82.9% |
|  | Micro | 86.8% | 86.8% | 86.8% |
| **Per-Class** | Functional/Psychogenic | 78.3% | 81.1% | 79.7% |
|  | Structural/Inflammatory | 70.8% | 67.2% | 69.0% |
|  | Healthy | 100% | 100% | 100% |

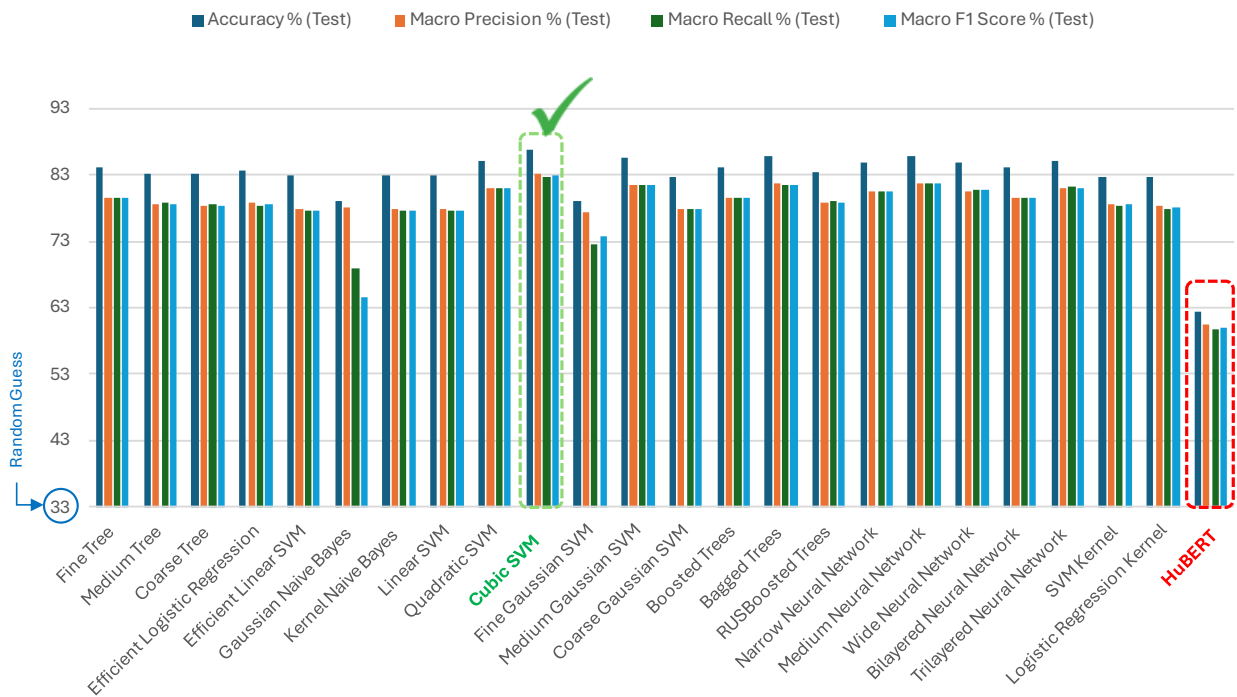

**Figure 5.** Test performance comparison across all evaluated models in Stage 2. The cubic SVM (Green) achieved the highest accuracy, outperforming all models including HuBERT (Red), and was selected as the best-performing classifier

Stage 2 delivered very strong separation between the three groups, reaching a Macro-averaged ROC-AUC of 0.965. This level of performance shows that the hierarchical structure is doing exactly what it was designed to do: break the problem down in a way that aligns with clinical logic.

The Healthy class reached a perfect ROC-AUC of 1.000, which is not surprising given the system's design. Because Stage 1 already does an excellent job filtering out Non-Pathological (Healthy) samples, Stage 2 can focus its efforts on the harder task of telling apart the two main pathological categories.

Both pathological groups were also well separated. The Functional/Psychogenic group achieved an ROC-AUC of 0.934, while the Structural/Inflammatory group reached 0.919. These results indicate that the model is consistently able to sort voices based on the underlying type of disorder. A full view of the multi-class performance, including the ROC and Precision-Recall curves, is shown in **Figure 6**. More detailed insights into where misclassifications occurred can be found in the confusion matrix in **Figure 7**.

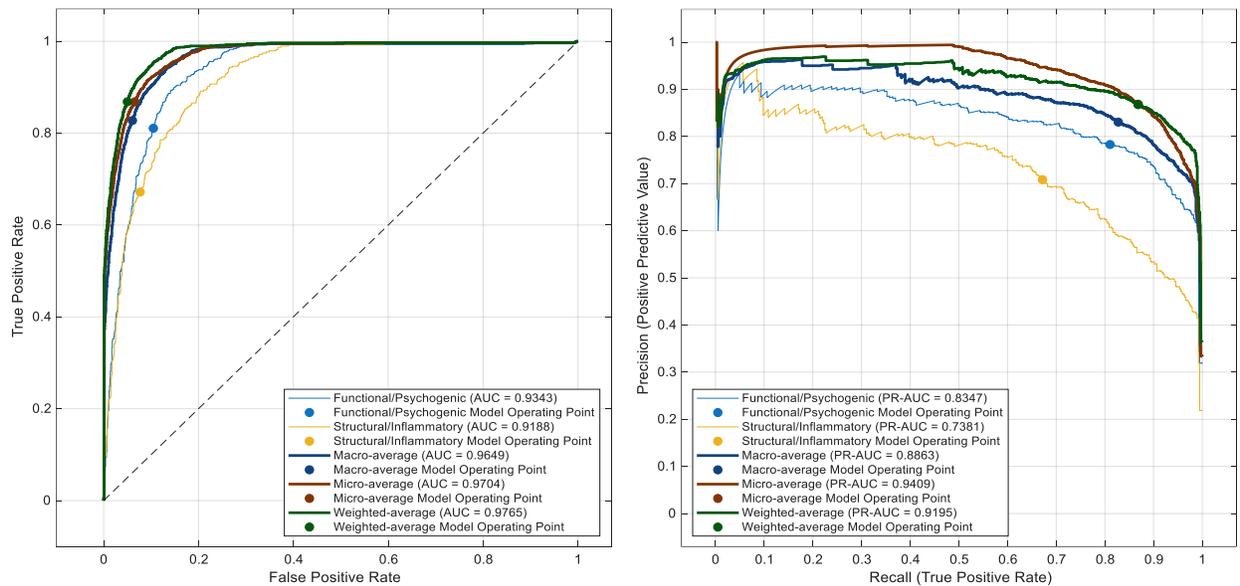

**Figure 6.** Multi-Class ROC and Precision-Recall Curves for Stage 2 Etiological Triage using the selected cubic SVM classifier. This figure presents the ROC and PR curves for the three-class classification task in Stage 2. Curves are shown separately for the Healthy, Functional/Psychogenic, and Structural/Inflammatory groups. The strong overall performance, reflected in the Macro-averaged ROC-AUC of 0.965, highlights how the hierarchical design supports clear separation of the etiological categories.

**Figure 7.** Confusion Matrix for Stage 2 Etiological Triage. This confusion matrix shows how well Stage 2 distinguishes among the three etiological groups. As expected, the Healthy class exhibits zero or very minimal confusion, since Stage 1's binary classifier effectively filters out Non-Pathological (Healthy) samples before they reach this stage.

### 3.3 Stage 3: Subtype Classification Performance

The final stage tackled the most challenging part of the pipeline: distinguishing among the nine specific voice disorder subtypes. The best results for this task came from a quadratic SVM (**Figure 8**), which used an expanded set of 25 features that included the predicted classes from both Stage 1 and Stage 2. These added features served as extra biomarkers that helped the model refine its decisions. The quadratic SVM models also outperformed the HuBERT model, which achieved an accuracy of 49.7%.

For a nine-class problem in which random guessing would yield only 11.1% accuracy, achieving a test accuracy of 73.7% is a strong outcome. Beyond accuracy, the model's ability to separate the classes remains very robust, as shown by a Macro-averaged ROC-AUC of 0.949. This indicates that the model consistently ranks the correct disorder class higher than the alternatives, which is particularly important when these predictions support clinical decision making.

One major factor influencing performance is the notable class imbalance in the Saarbruecken database. This imbalance becomes clear when comparing the aggregated performance metrics shown in **Table 4**. The sharp drop in the Macro F1-Score compared with the Micro F1-Score shows that while the model performs very well on the more common classes, it has more difficulty with the underrepresented ones. The macro metric treats all classes equally, so its lower value reflects this imbalance.

**Table 4**: Classification performance metrics for the nine-class subtype problem, highlighting the impact of class imbalance (Macro-Averaged versus Micro-Averaged scores).

| Metric | Macro-Averaged | Micro-Averaged | Weighted-Averaged |
|---|---|---|---|
| **Precision** | 55.20% | 73.30% | 72.70% |
| **Recall** | 46.50% | 73.30% | 73.30% |
| **F1-Score** | 45.00% | 73.30% | 69.70% |

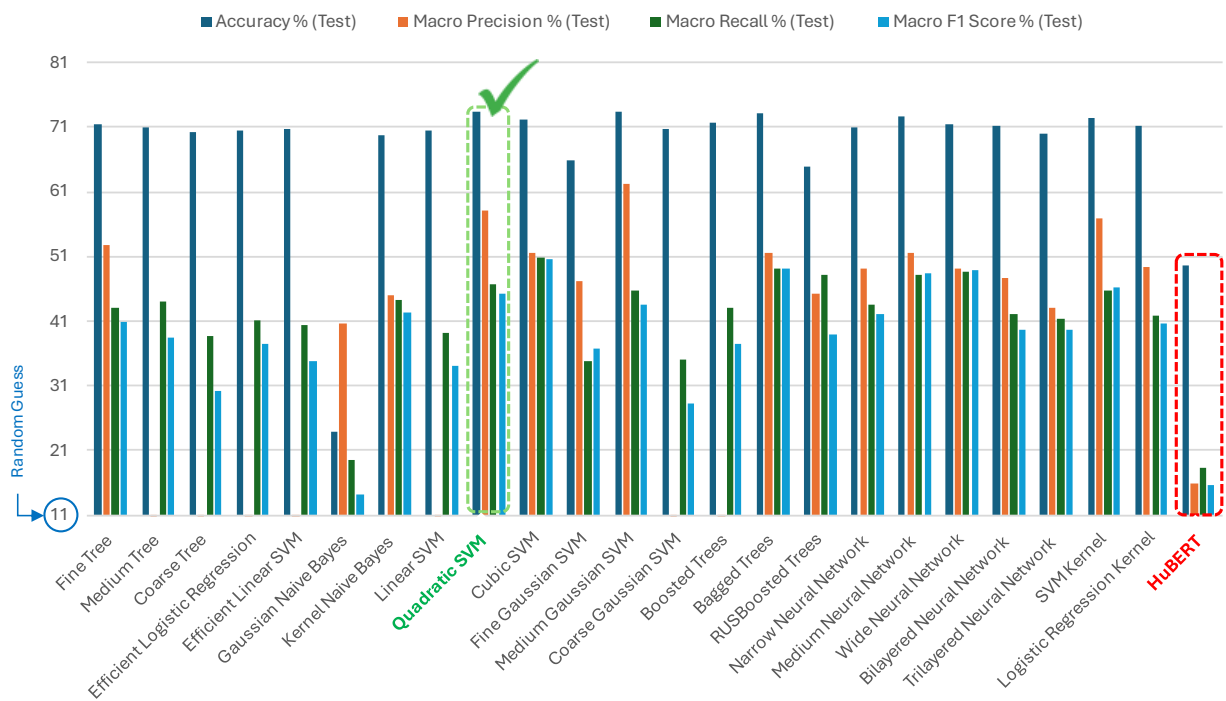

**Figure 8.** Test performance comparison across all evaluated models in Stage 3. The quadratic SVM (Green) achieved the highest accuracy, outperforming all models including HuBERT (Red), and was selected as the best-performing classifier

The Precision-Recall Area Under the Curve (PR-AUC) values further substantiate the observed performance trends, offering a crucial perspective on the model's performance on the imbalanced dataset. The highest PR-AUC scores are primarily concentrated within the Structural/Inflammatory disorder group, notably including Laryngitis (0.628), Reinke Edema (0.546), and Contact Pachydermia (0.540). These elevated scores are highly significant because they confirm that, for these specific conditions, the model is capable of achieving the necessary balance of both high precision and high recall simultaneously. This indicates a superior ability to correctly identify these disorders without generating excessive false alarms, making the classifier particularly trustworthy for these etiologies in a clinical setting. Conversely, the lower mean PR-AUC for the Functional/Psychogenic group (0.355) reflects the inherent difficulty in precisely isolating these classes due to their acoustic similarity and lower representation in the training data.

A closer inspection of the detailed class-level performance, summarized in **Table 5**, reveals a consistent pattern directly aligning with clinical knowledge. The Structural/Inflammatory disorders demonstrate consistently higher discriminability (Mean ROC-AUC: 0.961), making them substantially easier for the model to distinguish. In contrast, the Functional/Psychogenic disorders show lower performance metrics (Mean ROC-AUC: 0.903), indicating greater acoustic overlap and more variability in their vocal manifestation. This clear pattern has a rational clinical basis: physical lesions that characterize the Structural/Inflammatory group tend to induce more severe and acoustically consistent vocal fold perturbations, thus creating stronger, more reliable

"digital biomarkers" that are readily quantified by our derived acoustic features. This phenomenon is clearly reflected in the high individual ROC-AUC values observed for conditions like Reinke Edema and Contact Pachydermia.

**Table 5:** Detailed ROC-AUC and PR-AUC results for the eight specific disorder subtypes in Stage 3. The table highlights the superior discriminative performance for the Structural/Inflammatory group compared to the Functional/Psychogenic group, as shown by the mean AUC scores.

| Disorder Groups | Disorder | ROC-AUC | Mean ROC-AUC | PR-AUC | Mean PR-AUC |
|---|---|---|---|---|---|
| **Structural Inflammatory** | Contact Pachydermia | 0.97 | **0.96** | 0.53966 | **0.50** |
| | Laryngitis | 0.95 | | 0.62681 | |
| | Reinke Edema | 0.97 | | 0.54566 | |
| | Vocal Cord Polyp | 0.93 | | 0.30153 | |
| **Functional Psychogenic** | Functional Dysphonia | 0.89 | 0.90 | 0.33842 | 0.35 |
| | Hyperfunctional Dysphonia | 0.91 | | 0.50468 | |
| | Psychogenic Dysphonia | 0.90 | | 0.29304 | |
| | Dysodia | 0.91 | | 0.28165 | |
| **Average** | --- | **0.93** | --- | 0.4289 | --- |

The full multi-class performance is visualized in **Figure 9**, and the detailed classification results are presented in the nine-class confusion matrix, **Figure 10**.

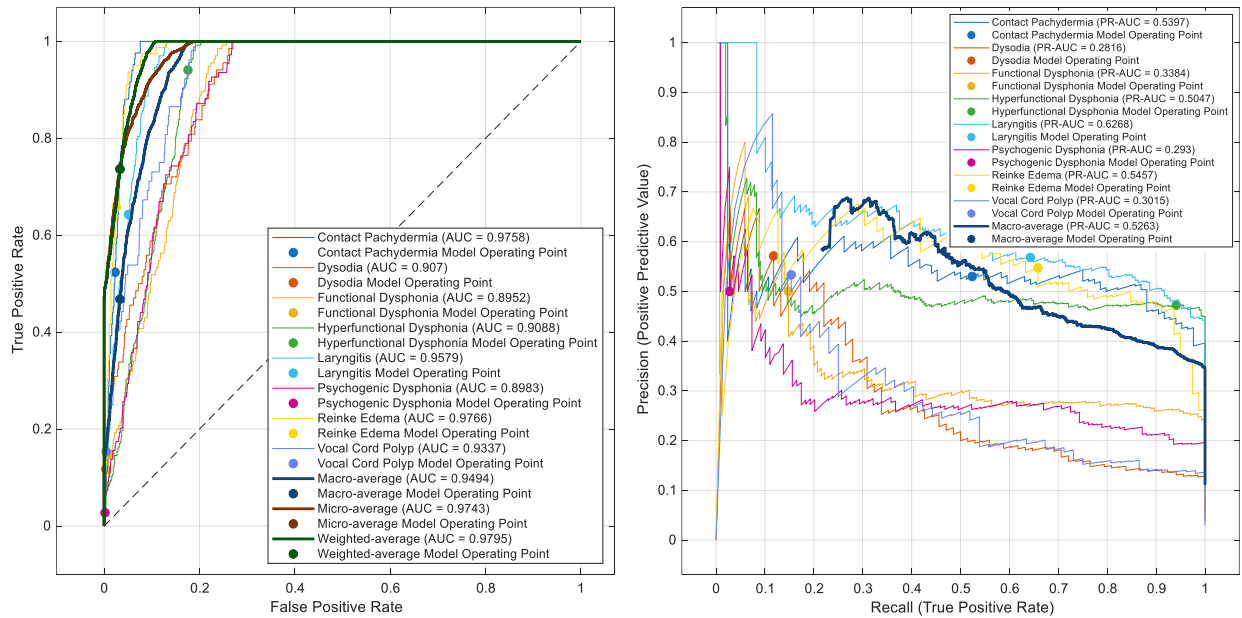

**Figure 9.** Multi-Class ROC and Precision-Recall Curves for Stage 3 Subtype Classification. This figure displays the ROC and PR curves for the 8-class classification task. The curves highlight the superior discriminability (high ROC-AUC) achieved, particularly for structural disorders, and the challenge (lower PR-AUC) in classifying the less-represented functional subtypes due to high class imbalance.

| True Class \ Predicted | Contact Pachydermia | Dysodia | Functional Dysphonia | Healthy | Hyperfunctional Dysphonia | Laryngitis | Psychogenic Dysphonia | Reinke Edema | Vocal Cord Polyp |
|---|---|---|---|---|---|---|---|---|---|
| Contact Pachydermia | 44 | | | 1 | | 36 | | | 4 |
| Dysodia | | 8 | 4 | | 56 | | | | |
| Functional Dysphonia | | 1 | 20 | | 112 | | 2 | | |
| Healthy | | | | 824 | | | | | |
| Hyperfunctional Dysphonia | | 2 | 12 | 1 | 239 | | 1 | | |
| Laryngitis | 29 | | | | | 108 | | 29 | 2 |
| Psychogenic Dysphonia | | 3 | 4 | | 99 | | 3 | | |
| Reinke Edema | 2 | | | 2 | | 24 | | 52 | 1 |
| Vocal Cord Polyp | 8 | | | 2 | | 22 | | 14 | 8 |

**Figure 10.** Confusion Matrix for Stage 3 Subtype Classification. The confusion matrix for Stage 3 illustrates the specific misclassification patterns among the 8 disorders and healthy control. It visually demonstrates the high true positive rates for the structural classes and the common confusion between related functional or psychogenic subtypes. As expected, the Healthy class exhibits zero confusion. This perfect separation is a direct result of the hierarchical architecture, where the Non-Pathological (Healthy) samples were already isolated by Stage 1 and 2 before reaching the subsequent refinement stages.

### 3.4 Baseline Comparison (Flat Classification)

To assess the advantage of the hierarchical structure, we trained an Ensemble Bagged Trees non-hierarchical (flat) classification model, which was the best-performing model among more than 20 candidates including HuBERT model (**Figure 11**), directly on the 21 acoustic features to classify all disorder classes simultaneously. The comparison between the hierarchical and the non-hierarchical, flat classification model unequivocally validates the benefits of incorporating a hierarchical approach for voice disorder diagnosis. While the flat model, which attempted to classify all disorder subtypes simultaneously, yielded a significantly lower Macro-averaged ROC-AUC of 0.632 (**Table 6**), the hierarchical framework achieved a high Average ROC-AUC of 0.9317 (**Table 5**). This substantial performance gap of 0.313 in Macro ROC-AUC demonstrates the efficacy of using sequential classification outputs from the higher stages as feature augmentation for subsequent, more granular classification tasks. The flat model struggled particularly with resolving ambiguity in disorders like Psychogenic Dysphonia (ROC-AUC: 0.5568) and Dysodia (ROC-AUC: 0.51649), where the hierarchical approach achieved high separability, reaching ROC-AUCs of 0.89829 and 0.90705, respectively.

This performance disparity extends to the Precision-Recall AUC (PR-AUC in **Table 6**), which is crucial for evaluating performance on imbalanced datasets. The flat baseline model demonstrated poor reliability, yielding a low Average PR-AUC of 0.142. In contrast, the hierarchical model achieved an Average PR-AUC of 0.4289, indicating a far superior ability to maintain higher precision while maximizing recall across all classes. Notably, the mean PR-AUC for Structural

Disorders drastically improved from 0.169 in the flat model to 0.5039 in the hierarchical model. This consistent and dramatic improvement in both ROC-AUC and PR-AUC metrics across all disorder groups confirms that breaking down the classification problem into hierarchical stages is a more robust and effective methodology than a single flat classification for complex voice disorder diagnosis.

**Table 6:** Detailed ROC-AUC and PR-AUC results for the eight specific disorder subtypes in Stage 3 but classified with the non-hierarchical (flat) model.

| Disorder Groups | Disorder | ROC-AUC | Mean ROC-AUC | PR-AUC | Mean PR-AUC |
|---|---|---|---|---|---|
| Structural Inflammatory | Contact Pachydermia | 0.68 | **0.69** | 0.13 | **0.17** |
| | Laryngitis | 0.71 | | 0.24 | |
| | Reinke Edema | 0.66 | | 0.14 | |
| | Vocal Cord Polyp | 0.71 | | 0.15 | |
| Functional Psychogenic | Functional Dysphonia | 0.60 | 0.57 | 0.10 | 0.11 |
| | Hyperfunctional Dysphonia | 0.62 | | 0.20 | |
| | Psychogenic Dysphonia | 0.55 | | 0.10 | |
| | Dysodia | 0.51 | | 0.03 | |
| **Average** | --- | **0.63** | --- | 0.14 | --- |

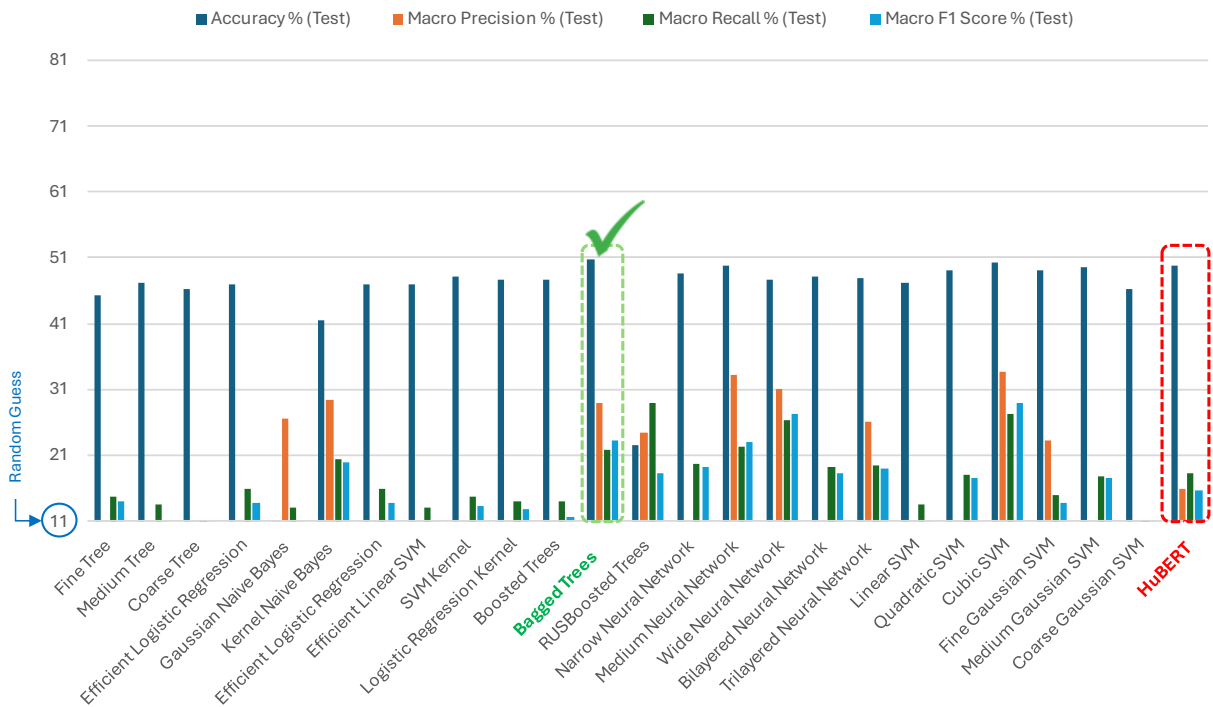

**Figure 11.** Test performance comparison across all evaluated models in Flat classifire. The Bagged Trees (Green) achieved the highest accuracy, outperforming all models including HuBERT (Red), and was selected as the best-performing classifier

## 4. Discussion

### 4.1 Efficacy of Hierarchical Feature Augmentation

The central outcome of this study is the strong and consistently validated performance of the proposed hierarchical classification framework for modeling benign laryngeal voice disorders. By design, the framework mimics the stepwise reasoning typically used in clinical decision-making. The system first establishes a reliable broad categorization, with Stage 1 achieving an accuracy of 80.5 percent and Macro-averaged ROC-AUC of 0.848 and PR-AUC of 0.830 and then uses this information to guide more detailed diagnostic stages (Stage 2). This staged architecture allows the downstream classifiers to operate within a narrower and more clinically meaningful context, effectively reducing the complexity of the subsequent tasks. As a result, broad pre-classification serves as a form of contextual feature enrichment, enabling the model to disentangle subtle acoustic differences that would be far more difficult to separate in a flat, nine-class setting. The final Stage 3 performance, reflected in a Macro-averaged ROC-AUC of 0.961, substantially surpasses what was achievable using a non-hierarchical approach (Macro-averaged ROC-AUC of 0.636) and highlights the advantages of this clinically inspired, sequential methodology.

These findings are also consistent with established knowledge of dysphonia physiology. Structural and inflammatory pathologies, such as Reinke's edema, Laryngitis or Contact Pachydermia, produce clear and measurable alterations in vocal fold mass, stiffness, and mucosal wave behavior. These physiological disturbances create distinctive acoustic patterns, often characterized by elevated shimmer, irregular vibratory cycles, and increased spectral noise. The combination of Mel-spectrogram representations with handcrafted acoustic features captures these patterns reliably, leading to stronger classification performance for these disorder categories. In contrast, functional and psychogenic disorders tend to arise from behavioral patterns, compensatory vocal techniques, or transient misuse rather than fixed anatomical changes. This inherent variability results in more diffuse and less predictable acoustic signatures, which naturally contributes to their lower PR-AUC values and the greater difficulty observed in differentiating these categories. Together, these results demonstrate not only the effectiveness of the hierarchical model but also its alignment with the underlying biomechanics and clinical presentation of voice disorders.

### 4.2 SVD Nomenclature Limitations for US Clinical Translation

While the SVD dataset provides a robust benchmark for voice disorder classification, its pathology nomenclature (e.g., "Dysodia") reflects European clinical conventions from the dataset's curation era and may not fully align with contemporary US laryngology terminology. This discrepancy could limit direct translatability to American practice settings and may contribute to the variable adoption of SVD-based models in US clinical workflows.

### 4.2 Underperformance of Flat, HuBERT and HeAR Classifiers

Our hierarchical model performs reliably across all stages and comparing it with standard baselines reveals clear shortcomings in both flat (non-hierarchical) classifiers and general pre-trained large speech models such as HuBERT and HeAR. These models struggle with the subtle nature of benign laryngeal disorders extracted from short, sustained vowel recordings. Their performance highlights why clinically informed, domain-specific frameworks are necessary for this type of task.

- **Limitations of Flat Classification**
  The flat Ensemble Bagged Trees model was trained directly on the 21 handcrafted acoustic features to classify all nine categories in a single step. This reflects a typical one-stage design without any triage structure. Although we evaluated more than 20 candidate models (**Figure 11**), this approach achieved only a macro-ROC-AUC of 0.632 (**Table 6**). Its main weaknesses appeared in groups with similar acoustic patterns. Psychogenic Dysphonia (ROC-AUC: 0.557) and Functional Dysphonia (ROC-AUC: 0.597) share overlapping characteristics in common features such as jitter which often led to confusion with structural lesions or even healthy voices.
  Because the model attempts to separate all classes at once, without the benefit of early-stage binary screening or later grouping by etiology, it becomes overloaded by imbalanced data and mixed severity levels. This directly contributes to a ~23% drop in accuracy (50.7% compared with our 73.66%) and a 0.317 reduction in Macro-averaged ROC-AUC (0.632 compared with 0.9494). Flat models may be efficient, but they do not reflect how clinicians' reason through differential diagnosis. As a result, they increase the risk of inconsistent automated predictions and limit practical scalability for triage.

- **Limitations of HuBERT**
  The results from our HuBERT model reinforce these observations. Since HuBERT is trained mainly on conversational speech, such as LibriSpeech, its representations capture natural prosody well but are not tuned for short, controlled sustained vowels. These vowels contain precise dysphonic cues such as breathy noise or subharmonics that HuBERT fails to detect effectively, even after fine-tuning with fixed-length padding at 3 seconds and 16 kHz resampling.
  This domain mismatch resulted in accuracy deficits of 5 to 25 percent across the stages. For example, Stage 3 accuracy dropped to 49.72%, while our model reached 73.66%. In contrast, our fusion approach, which combines Mel-spectrogram CNNs with targeted acoustic features such as shimmer and harmonic-to-noise ratio, captures perturbations directly related to vocal fold biomechanics. This leads to sharper class separation.

- **Limitations of HeAR**
  HeAR is not well suited for sustained vowel analysis because it is trained on short, event-driven health sounds such as coughs and breathing, rather than quasi-stationary phonation. Its masked autoencoder objective emphasizes global spectrotemporal structure over the fine-grained, cycle-to-cycle variations in fundamental frequency, harmonic stability, and noise that characterize sustained vowels. Consistent with this limitation, HeAR exhibited a marked reduction in performance in Stage 1 binary classification (58.43% accuracy); therefore, it was not included in the multiclass classification experiments of Stages 2 and 3.

Overall, both flat classification and HuBERT models illustrate the limitations of generic or non-specialized designs for voice disorder triage. Flat models lack the stepwise clinical reasoning structure, and HuBERT struggles with the domain shift from conversational speech to sustained clinical phonation. Similarly, HeAR, while effective for transient health acoustic events, is not optimized for quasi-stationary vowel production and underrepresents fine-grained phonatory irregularities, limiting its utility for dysphonia assessment. Our hierarchical and biomarker-guided design overcomes these issues by combining interpretable acoustic markers with deep spectral learning, providing a more explainable and clinically aligned solution for assessing laryngeal disorders.

### 4.3 Voice Biomarkers as Foundations for Digital Health

The features employed in this framework are not merely mathematical constructs but quantitative voice biomarkers that reflect the underlying biomechanical state of the vocal folds and respiratory system. The high discriminative power achieved throughout the framework confirms the potential of automated sustained phonation analysis as a highly scalable and non-invasive diagnostic tool for digital health.

Integration of this technology into mobile or web-based applications offers a significant opportunity for decentralized clinical deployment:

- **Mass Screening and Early Detection:** Stage 1's high sensitivity can be employed for proactive monitoring in high-risk populations (e.g., professional voice users, patients with chronic diseases like Parkinson's, or post-laryngeal surgery follow-up), facilitating immediate referral and reducing critical diagnostic delays.
- **Clinical Triage Guidance:** Stage 2's etiological classification provides immediate guidance for clinical triage, distinguishing patients likely needing medical or surgical intervention (structural and inflammatory) from those best treated by behavioral voice therapy (functional and psychogenic).
- **Longitudinal** Monitoring: The fine-grained Stage 3 output can serve as a sensitive, objective metric for tracking disease progression, evaluating the efficacy of therapeutic interventions, and monitoring for relapse in chronic conditions.

By transforming acoustic signals into actionable digital biomarkers, this system provides a pathway toward reducing the substantial diagnostic burden and associated clinical costs of dysphonia and in general voice disorders, furthering the goal of personalized and preemptive digital healthcare.

### 4.4 Theoretical Multi-Vowel Fusion Strategies

Although the present analysis focuses on single vowel classification, real clinical use requires decisions at the level of the individual patient rather than at the level of isolated vocalizations. Each participant in the dataset contributes twelve recordings that include three sustained vowels (/a/, /i/, /u/) produced under four phonatory conditions: neutral pitch, high pitch, low pitch, and a continuous low-high-low glide. This collection of repeated vocal samples offers a valuable opportunity to strengthen diagnostic confidence by integrating information across several observations. For this reason, the development of a multi vowel fusion strategy represents an important direction for improving subject level diagnostic performance.

If each vowel is considered an independent observation with a per vowel accuracy of *p*, the expected accuracy of a subject level decision based on *k* repeated vowels can be estimated through a binomial majority vote formulation:

$$Acc_{Subject}(k) = \sum_{i=[k/2]}^{k} \binom{k}{i} p^i (1-p)^{k-i} \qquad (5)$$

Using the empirically measured single vowel accuracy from Stage 1, which is p=0.805, this model predicts a meaningful improvement when multiple observations are fused. For example, combining five vowels yields an estimated subject level accuracy of approximately 0.946, and using eleven vowels increases this estimate to about 0.990. These values are theoretical, but they clearly demonstrate the potential benefits of aggregating several vocal samples from each patient to reduce uncertainty and increase the stability of the classification outcome.

These theoretical observations highlight the importance of pursuing and validating multi vowel fusion methods in future investigations. Clinical voice examinations routinely include repeated phonations, and a subject level fusion approach can take advantage of this natural redundancy. Incorporating this additional information has the potential to greatly enhance the overall robustness and reliability of the system as a digital diagnostic biomarker.

## 5. Conclusion

This work developed and validated a hierarchical machine learning framework inspired by clinical decision-making to classify benign laryngeal voice disorders. The system follows a structured, triage-like pathway in which voice recordings are first screened in a binary manner, then grouped into broader etiological categories, and finally assigned to specific disorder subtypes. By combining deep spectral features with interpretable acoustic measures, the proposed framework outperformed traditional flat machine learning–based classifiers as well as pre-trained large speech and audio models, including HuBERT and HeAR, whose generic pre-training objectives are not optimized for sustained clinical phonation.

In the final diagnostic stage, the model achieved a macro-averaged ROC–AUC of approximately 0.95, demonstrating that short, sustained vowels contain reliable quantitative information for distinguishing among diverse structural and inflammatory laryngeal conditions. Beyond its strong diagnostic performance, the framework is designed as a scalable and non-invasive solution that can be integrated into digital health platforms. Its potential applications include large-scale early screening, accelerated clinical assessment, and personalized longitudinal monitoring of vocal health, representing a meaningful step toward reducing the global burden of dysphonia.

### *Conflicts of Interest*

Anaïs Rameau owns equity in Perceptron Health, Inc. and serves as an advisor for Pentax Medical and Sound Health Systems, Inc. The other authors have no conflicts of interest to disclose.

*Acknowledgments*

Anaïs Rameau was supported by a Paul B. Beeson Emerging Leaders Career Development Award in Aging (K76 AG079040) from the National Institute on Aging and by the Bridge2AI award (OT2 OD032720) from the NIH Common Fund.

**References**

[1] C. Gobl and A. N. Chasaide, "The role of voice quality in communicating emotion, mood and attitude," *Speech communication,* vol. 40, no. 1-2, pp. 189-212, 2003.

[2] I. Grichkovtsova, M. Morel, and A. Lacheret, "The role of voice quality and prosodic contour in affective speech perception," *Speech Communication,* vol. 54, no. 3, pp. 414-429, 2012.

[3] S. Aghadoost, S. Jalaie, P. Dabirmoghaddam, and S. M. Khoddami, "Effect of muscle tension dysphonia on self-perceived voice handicap and multiparametric measurement and their relation in female teachers," *Journal of Voice,* vol. 36, no. 1, pp. 68-75, 2022.

[4] B. Ebersole, R. S. Soni, K. Moran, M. Lango, K. Devarajan, and N. Jamal, "The influence of occupation on self-perceived vocal problems in patients with voice complaints," *Journal of Voice,* vol. 32, no. 6, pp. 673-680, 2018.

[5] M. STERCZEWSKI, D. TRZASKOMA, and K. SZKLANNY, "The Effect of Vocal Effort on Voice Quality in Occupations with High Vocal Load," *Engineering Transactions,* 2025.

[6] L. Craig, J. Cameron, and E. Longden, "Work-related experiences of people who hear voices: An occupational perspective," *British journal of occupational therapy,* vol. 80, no. 12, pp. 707-716, 2017.

[7] A. L. Spina, R. Maunsell, K. Sandalo, R. Gusmão, and A. Crespo, "Correlation between voice and life quality and occupation," *Revista Brasileira de Otorrinolaringologia,* vol. 75, pp. 275-279, 2009.

[8] S. M. Khoddami, S. Aghadoost, A. R. Khatoonabadi, P. Dabirmoghaddam, and S. Jalaie, "Comparison and Relation Between Vocal Tract Discomfort and Voice Handicap Index in Teachers With and Without Muscle Tension Dysphonia," *Archives of Rehabilitation,* vol. 24, no. 2, pp. 264-283, 2023.

[9] L. Allen and A. Hu, "Voice disorders in the workplace: a scoping review," *Journal of Voice,* vol. 38, no. 5, pp. 1156-1164, 2024.

[10] R. B. Fujiki and S. L. Thibeault, "Voice disorder prevalence and vocal health characteristics in adolescents," *JAMA Otolaryngology–Head & Neck Surgery,* vol. 150, no. 9, pp. 800-810, 2024.

[11] C. Sapienza and B. Hoffman, *Voice disorders*. Plural Publishing, 2020.

[12] R. H. G. Martins, H. A. do Amaral, E. L. M. Tavares, M. G. Martins, T. M. Gonçalves, and N. H. Dias, "Voice disorders: etiology and diagnosis," *Journal of voice,* vol. 30, no. 6, pp. 761. e1-761. e9, 2016.

[13] H. Byeon, "The risk factors related to voice disorder in teachers: a systematic review and meta-analysis," *International journal of environmental research and public health,* vol. 16, no. 19, p. 3675, 2019.

[14] S. R. Schwartz *et al.*, "Clinical practice guideline: hoarseness (dysphonia)," *Otolaryngology–Head and Neck Surgery,* vol. 141, no. 1_suppl, pp. 1-31, 2009.

[15] W. Angerstein *et al.*, "Diagnosis and differential diagnosis of voice disorders," in *Phoniatrics I: Fundamentals–Voice Disorders–Disorders of Language and Hearing Development*: Springer, 2019, pp. 349-430.

[16] O. Aghadoost, N. Moradi, A. Aghadoost, A. Montazeri, M. Soltani, and A. Saffari, "A comparative study of Iranian female primary school teachers' quality of life with and without voice complaints," *Journal of Voice,* vol. 30, no. 6, pp. 688-692, 2016.

[17] S. Krischke *et al.*, "Quality of life in dysphonic patients," *Journal of Voice,* vol. 19, no. 1, pp. 132-137, 2005.

[18] M. D. Morrison, H. Nichol, and L. Rammage, *The management of voice disorders*. Springer, 2013.

[19] R. T. Sataloff, *Treatment of voice disorders*. Plural Publishing, 2017.

[20] R. R. Patel *et al.*, "Recommended protocols for instrumental assessment of voice: American Speech-Language-Hearing Association expert panel to develop a protocol for instrumental assessment of vocal function," *American journal of speech-language pathology,* vol. 27, no. 3, pp. 887-905, 2018.

[21] O. Aghadoost, Y. Amiri-Shavaki, N. Moradi, and S. Jalai, "A comparison of dysphonia severity index in female teachers with and without voice complaints in elementary schools of Tehran, Iran," *Nurs Midwifery Stud,* vol. 1, no. 3, pp. 133-8, 2013.

[22] S. Aghadoost, Y. Molazeinal, S. M. Khoddami, G. Shokuhifar, P. Dabirmoghaddam, and M. Saffari, "Dysphonia severity index and consensus auditory-perceptual evaluation of voice outcomes, and their relation in hospitalized patients with COVID-19," *Journal of Voice,* vol. 39, no. 3, pp. 853. e1-853. e8, 2025.

[23] M. Brockmann-Bauser, "Instrumental analysis of voice," *The Handbook of Clinical Linguistics, Second Edition,* pp. 523-537, 2024.

[24] M. Fantini *et al.*, "The rapidly evolving scenario of acoustic voice analysis in otolaryngology," *Cureus,* vol. 16, no. 11, 2024.

[25] V. Parsa and D. G. Jamieson, "Acoustic discrimination of pathological voice," *Journal of Speech, Language, and Hearing Research,* vol. 44, no. 2, pp. 327-339, 2001.

[26] S.-S. Wang, C.-T. Wang, C.-C. Lai, Y. Tsao, and S.-H. Fang, "Continuous Speech for Improved Learning Pathological Voice Disorders," *IEEE Open Journal of Engineering in Medicine and Biology,* vol. 3, pp. 25 - 33, 2022.

[27] L. Verde, N. Brancati, G. D. Pietro, M. Frucci, and G. Sannino, "A Deep Learning Approach for Voice Disorder Detection for Smart Connected Living Environments," *ACM Trans. Internet Technol.,* vol. 22, no. 1, p. Article 8, 2021, doi: 10.1145/3433993.

[28] G. Brindha, P. S, S. N. N, V. K. G, and V. K. K, "Detection of Voice Pathologies and Classification using Deep Learning in Healthcare," *2025 6th International Conference on Inventive Research in Computing Applications (ICIRCA),* pp. 1945-1950, 2025.

[29] R. Islam, E. Abdel-Raheem, and M. Tarique, "Voice pathology detection using convolutional neural networks with electroglottographic (EGG) and speech signals," *Computer Methods and Programs in Biomedicine Update,* vol. 2, p. 100074, 2022/01/01/ 2022, doi: https://doi.org/10.1016/j.cmpbup.2022.100074.

[30] M.Pützer and W.J.Barry. *Saarbruecken Voice Databas*, doi: 10.5281/zenodo.16874898.

[31] A. Awad, M. A. A. Eldosoky, A. M. Soliman, and N. M. Mahmoud, "Automatic diagnosis of hyperkinetic dysphonia from speech recordings based on deep learning approaches," *Engineering Research Express,* vol. 7, no. 3, p. 035263, 2025/08/20 2025, doi: 10.1088/2631-8695/adf9c3.

[32] A. Roitman *et al.*, "Harnessing machine learning in diagnosing complex hoarseness cases," (in eng), *Am J Otolaryngol,* vol. 46, no. 1, p. 104533, Jan-Feb 2025, doi: 10.1016/j.amjoto.2024.104533.

[33] J. Reid, P. Parmar, T. Lund, D. K. Aalto, and C. C. Jeffery, "Development of a machine-learning based voice disorder screening tool," *American Journal of Otolaryngology,* vol. 43, no. 2, p. 103327, 2022/03/01/ 2022, doi: https://doi.org/10.1016/j.amjoto.2021.103327.

[34] F. T. Al-Dhief *et al.*, "Voice Pathology Detection and Classification by Adopting Online Sequential Extreme Learning Machine," *IEEE Access,* vol. 9, pp. 77293-77306, 2021, doi: 10.1109/ACCESS.2021.3082565.

[35] J.-Y. Lee, "Experimental Evaluation of Deep Learning Methods for an Intelligent Pathological Voice Detection System Using the Saarbruecken Voice Database," *Applied Sciences,* vol. 11, no. 15, p. 7149, 2021. [Online]. Available: https://www.mdpi.com/2076-3417/11/15/7149.

[36] S. Coelho and H. L. Shashirekha, "Identification of Voice Disorders: A Comparative Study of Machine Learning Algorithms," presented at the Speech and Computer: 25th International Conference, SPECOM 2023, Dharwad, India, November 29 – December 2, 2023, Proceedings, Part I, Dharwad, India, 2023. [Online]. Available: https://doi.org/10.1007/978-3-031-48309-7_45.

[37] J. Vrba *et al.*, "Reproducible Machine Learning-Based Voice Pathology Detection: Introducing the Pitch Difference Feature," *Journal of Voice*, doi: 10.1016/j.jvoice.2025.03.028.

[38] P. Gulsen, A. Gulsen, and M. Alci, "Machine Learning Models With Hyperparameter Optimization for Voice Pathology Classification on Saarbrücken Voice Database," *Journal of Voice,* 2025/01/07/ 2025, doi: https://doi.org/10.1016/j.jvoice.2024.12.009.

[39] W.-N. Hsu, B. Bolte, Y.-H. H. Tsai, K. Lakhotia, R. Salakhutdinov, and A. Mohamed, "Hubert: Self-supervised speech representation learning by masked prediction of hidden units," *IEEE/ACM transactions on audio, speech, and language processing,* vol. 29, pp. 3451-3460, 2021.

[40] S. Baur *et al.*, "HeAR--Health Acoustic Representations," *arXiv preprint arXiv:2403.02522,* 2024.